\documentclass[showpacs]{revtex4}
\usepackage{epsfig}
\def\beq{\begin{equation}}
\def\eeq{\end{equation}}
\def\beqa{\begin{eqnarray}}
\def\eeqa{\end{eqnarray}}

\def\GeV{\nobreak\,\mbox{GeV}}

\def\pli{p^\prime}
\def\ql{{p^\prime}^2}

\def\muli{\mu^\prime}
\def\nuli{\nu^\prime}
\def\ali{\alpha^\prime}

\begin{document}
\title{$\rho D^*D^*$ vertex from QCD sum rules}
\author{M.E. Bracco, M. Chiapparini}
\affiliation{Instituto de F\'{\i}sica, Universidade do Estado do Rio de 
Janeiro, 
Rua S\~ao Francisco Xavier 524, 20550-900 Rio de Janeiro, RJ, Brazil}
\author{F.S. Navarra and M. Nielsen}
\affiliation{Instituto de F\'{\i}sica, Universidade de S\~{a}o Paulo, 
C.P. 66318, 05389-970 S\~{a}o Paulo, SP, Brazil.}

\begin{abstract}
We calculate the  form factors  and the coupling constant in  the 
$\rho D^* D^*$ vertex in the framework of QCD sum rules. We evaluate  the 
three point correlation functions of the vertex considering both $\rho$ and $D^*$ 
mesons off--shell. The form factors obtained are very different but  give 
the same coupling constant: $g_{\rho D^* D^*} = 6.6 \pm 0.31$. This number is 50 \% 
larger than what we would expect from SU(4) estimates. 
\end{abstract}


\maketitle

\section{Introduction}

Charmonium production is a very useful source of information in heavy ion 
collisions. The knowledge of the   $J/\psi$ production rate can improve our
understanding  of these collisions and help us to know 
if there was a ``color glass condensate''  in the initial state. 
Charmonium production is very sensitive to  the 
existence and to the properties of the intermediate ``quark gluon plasma'' \cite{larry}.  
All the 
interesting effects happening in  these initial and intermediate phases can be blurred by 
interactions in the final stage of these collisions, when charmonium states interact with 
other comovers such as pions, $\rho$ mesons and nucleons, which form a hot hadronic gas. 
Since these interactions occur at an energy of the order of magnitude of the temperature, 
($\simeq 100 - 150$ MeV),  their  study  has to be made with 
non-perturbative methods. These can be  QCD sum rules \cite{svz},  
quark models and  the effective Lagrangian approach \cite{linko,su}.
This last approach  has been developed for almost ten years now and a 
great progress in the 
understanding of the interactions of charmed mesons with light mesons and nucleons has been 
achieved. Part of this progress is due to a persistent study of the vertices involving 
charmed mesons, namely $D^* D \pi$ \cite{nnbcs00,nnb02}, 
$D D \rho$\cite{bclnn01},  $D D J/\psi$ \cite{mnns02},  $D^* D J/\psi$ \cite{smnn04,mnns05},  
$D^* D^* \pi$ \cite{wang,cdnn05}, $D^* D^* J/\psi$ \cite{bcnn05},   
$D_s D^* K$, $D_s^* D K$ \cite{bccln06} and  $D D \omega$  \cite{hmm07}. 
More specifically,   it is very important to know the 
precise functional form of the form factors in these vertices and even to know how this form 
changes when one or the other (or both) mesons are off-shell. This careful determination of 
the  charm form factors has been done bit by bit over the last seven years in the framework 
of QCD sum rules, which are the best tool to give a first principles answer to this problem. 

Understanding charmonium production in heavy ion collisions would be  already a good reason  
to  study of hadronic charm form factors. However, since 2003, 
due the precise measurements of $B$ decays  performed by BELLE, BES and BABAR, this subject 
gained a new relevance. In $B$ decays new particles have been observed, such as the 
$D_{sJ} (2317)$ and the $X(3872)$. These particles very often decay into  an intermediate two 
body state, which then undegoes final state interactions, with the exchange of one or more 
virtual mesons.  As an example of  specific situation where a precise knowledge of the 
$ \rho  D^*  D^*  $ form factor is required, we may consider the decay 
$X(3872) \, \rightarrow \, J/\psi \, + \, \rho$. As suggested in \cite{lzz}, this decay 
proceeds in two steps. First the $X$ decays into a $D$ - $D^*$ intermediate state and then
these two particles exchange a $D^*$ producing the final $J/\psi$ and $\rho$. This is shown  
in Fig. 1b and 1f of   \cite{lzz}. In order to compute the effect of these interactions 
in the final decay rate we need the $ \rho  D^*  D^*  $ form factor. 

In the present paper we 
calculate this form factor with QCDSR. The  $ \rho  D^*  D^*  $ vertex is similar to
the  $J/\psi   D^*  D^*  $ vertex treated in \cite{bcnn05}. 
As  before, because there are three vector particles involved, the
number of Lorentz structures is very large and we have to choose a reliable one to 
perform the calculations.  Here we introduce the pole-continuum analysis and impose the
pole dominance as a criterion to reduce the freedom in the choice  of the Borel parameter.
In the next section, for completeness we describe the QCDSR technique and in 
section III we present the  results and compare them with results obtained in other works.  

\section{The sum rule for the  $ \rho  D^*  D^*   $  vertex}

Following our previous works and especially Ref. 
\cite{bcnn05}, we write the three-point function associated 
with the $\rho D^*D^*$ vertex, which is given by
\begin{equation}
\Gamma_{\nu \alpha \mu}^{(\rho)}(p,\pli)=\int d^4x \, d^4y \;\;
e^{i\pli\cdot x} \, e^{-i(\pli-p)\cdot y}
\langle 0|T\{j_{\mu}^{{{D^{*0}}} }(x) j_{\alpha}^{\rho^+ \dagger}(y) 
 j_{\nu}^{{D^{*-}} \dagger}(0)\}|0\rangle\, \label{correrhooff} 
\end{equation}
for an off-shell $\rho^+ $ meson, and:
\begin{equation}
\Gamma_{\nu \alpha \mu}^{({D^{*-}})}(p,\pli)=\int d^4x \, 
d^4y \;\; e^{i\pli\cdot x} \, e^{-i(\pli-p)\cdot y}\;
\langle 0|T\{j_{\nu}^{{D^{*0}}}(x)  j_{\alpha}^{{D^{*-}} \dagger}(y) 
 j_{\alpha}^{\rho^+ \dagger}(0)\}|0\rangle\, ,\label{corredmenosoff} 
\end{equation}
for an off-shell ${D^{*-}}$ meson. The general expression for the vertices 
(\ref{correrhooff}) and (\ref{corredmenosoff}) 
has fourteen independent Lorentz structures. We can write each 
$\Gamma_{\nu \alpha \mu}$ in terms of the invariant amplitudes associated 
with each one of these structures in the following form:
\begin{eqnarray}
\Gamma_{\mu\nu\alpha}(p,\pli)&=&
    \Gamma_1(p^2 , \ql , q^2) g_{\mu \nu} p_{\alpha} 
  + \Gamma_2(p^2,\ql, q^2) g_{\mu \alpha} p_{\nu} 
  + \Gamma_3(p^2,\ql , q^2) g_{\nu \alpha} p_{\mu} 
  + \Gamma_4(p^2,\ql ,q^2) g_{\mu \nu} \pli_{\alpha} \nonumber \\ 
&&+ \Gamma_5(p^2, \ql ,q^2) g_{\mu \alpha} \pli_{\nu} 
  + \Gamma_6(p^2,\ql ,q^2) g_{\nu\alpha} \pli_{\mu}  
  + \Gamma_7(p^2,\ql ,q^2) p_{\mu} p_{\nu} p_{\alpha}
  + \Gamma_8(p^2,\ql ,q^2) \pli_{\mu} p_{\nu} p_{\alpha}  \nonumber \\
&&+ \Gamma_9(p^2,\ql ,q^2) p_{\mu} \pli_{\nu} p_{\alpha}  
  + \Gamma_{10}(p^2,\ql ,q^2) p_{\mu} p_{\nu} \pli_{\alpha} 
  + \Gamma_{11}(p^2,\ql ,q^2) \pli_{\mu} \pli_{\nu} p_{\alpha}
  + \Gamma_{12}(p^2,\ql ,q^2) \pli_{\mu} p_{\nu} \pli_{\alpha}  \nonumber \\
&&+ \Gamma_{13}(p^2,\ql ,q^2) p_{\mu} \pli_{\nu} \pli_{\alpha} 
  + \Gamma_{14}(p^2,\ql ,q^2) \pli_{\mu} \pli_{\nu} \pli_{\alpha}
  \label{trace}  
\end{eqnarray}

Equations~(\ref{correrhooff}) and 
(\ref{corredmenosoff}) can be calculated in two diferent ways: using quark 
degrees of freedom --the theoretical or QCD side-- or using hadronic 
degrees of freedom --the phenomenological side. In the QCD side the 
correlators are evaluated  using the 
Wilson operator product expansion (OPE). The OPE incorporates the effects 
of the QCD vacuum through an infinite series of condensates of in\-crea\-sing 
dimension. On the other hand, the representation in terms of 
hadronic degrees of freedom is responsible for the introduction of the form 
factors, decay constants and masses. Both representations are matched invoking 
the quark-hadron global duality.
\subsection{The OPE side}

In the OPE or theoretical side  each meson interpolating
field appearing in Eqs.~(\ref{correrhooff}) and (\ref{corredmenosoff}) can be written 
in terms of the quark field operators in the following form: 
\beq
j_{\nu}^{\rho^+}(x) = \bar d(x) \gamma_{\nu} u(x)
\label{corho}
\eeq 
and
\beq
j_{\mu}^{D^{*-}}(x) = \bar c(x) \gamma_{\mu} d(x) 
\label{cods}
\eeq
where 
$u$, $d$ and $c$ are the up, down and charm quark field respectively. Each 
one of these currents has the same quantum numbers of the associated
meson. 
 
\begin{figure}[h]
\begin{picture}(12,3.5)
\put(0.0,0.5){\vector(1,0){1.5}}
\put(3.5,0.5){\vector(-1,0){2}}
\put(3.5,0.5){\vector(1,0){1.5}}
\put(1.5,0.5){\vector(1,1){1}}
\put(2.5,1.5){\vector(1,-1){1}}
\put(2.5,3){\vector(0,-1){1.5}}
\put(2.65,2.75){$q_\alpha$}
\put(0.25,0.65){$p_\mu$}
\put(4.55,0.65){$p'_\nu$}
\put(2.4,0.2){$\bar c$}
\put(1.85,1.1){$d$}
\put(3,1.1){$u$}
\put(2.4,1.2){$y$}
\put(1.75,0.53){$0$}
\put(3.05,0.53){$x$}
\put(2.0,2.2){$\rho^+$}
\put(0.35,0.1){$\bar{D^{*-}}$}
\put(4,0.1){$\bar{D^{*0}}$}
\put(7,0.5){\vector(1,0){1.5}}
\put(10.5,0.5){\vector(-1,0){2}}
\put(10.5,0.5){\vector(1,0){1.5}}
\put(8.5,0.5){\vector(1,1){1}}
\put(9.5,1.5){\vector(1,-1){1}}
\put(9.5,3){\vector(0,-1){1.5}}
\put(9.65,2.75){$q_\alpha$}
\put(7.25,0.65){$p_\mu$}
\put(11.55,0.65){$p'_\nu$}
\put(9.4,0.2){$\bar u$}
\put(8.85,1.1){$d$}
\put(10,1.1){$c$}
\put(9.4,1.2){$y$}
\put(8.75,0.53){$0$}
\put(10.05,0.53){$x$}
\put(8.75,2.2){$\bar{D^{*-}}$}
\put(7.35,0.1){$\rho^+$}
\put(11,0.1){$\bar{D^{*-}}$}
\end{picture}
\caption{Perturbative diagrams for the $\rho$ off-shell (left) and $D^*$
off-shell (right) correlators.}
\label{fig1}
\end{figure}
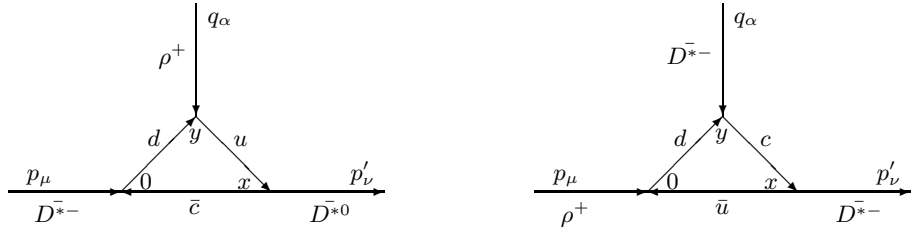

For each one of the invariant amplitudes appearing in Eq.(\ref{trace}), we 
can write a double dispersion relation over the virtualities $p^2$ and 
${\pli}^2$, holding $Q^2= -q^2$ fixed:
\begin{equation}
\Gamma_i(p^2,{\pli}^2,Q^2)=-\frac{1}{\pi^2}\int_{s_{min}}^\infty ds
\int_{u_{min}}^\infty du \:\frac{\rho_i(s,u,Q^2)}{(s-p^2)(u-{\pli}^2)}\;,
\;\;\;\;\;\;i=1,\ldots,14 \label{dis}
\end{equation}
where $\rho_i(s,u,Q^2)$ equals the double discontinuity of the amplitude
$\Gamma_i(p^2,{\pli}^2,Q^2)$, calculated using the Cutkosky's rules. 
The invariant amplitudes receive contributions 
from all terms in the OPE. The first one of those contributions comes from 
the perturbative term and it is represented in Fig.~\ref{fig1}.

We can work with any structure appearing in
Eq.(\ref{trace}), but we must choose those which have less ambiguities in 
the QCD sum rules approach, which means, less influence from the higher dimension 
condensates  and a better stability as a function of the Borel mass.
We have chosen the $g_{\mu \alpha} q_{\nu}$  structure.
In this structure the quark condensate (the condensate of lower 
dimension) contributes in the case of $D^*$ meson off-shell. 

 The corresponding perturbative spectral densities which enter in 
Eq.~(\ref{dis}) are

\begin{equation}
\rho^{(\rho)}(s,u,Q^2)=\frac{3}{2\pi\sqrt\lambda}
\left[\left(\frac{s-u-t}{2}-(2m_c^2)\right)\left(\frac{B-A}{2}\right) 
+ 2(J-I)+\frac{\pi}{2}(2m_c^2-u-s) -  D \right] \label{ddrhoff}
\end{equation}
for $\rho$ off-shell, and
\begin{equation}
\rho^{(D^*)}(s,u,Q^2)=-\frac{3}{2\pi\sqrt\lambda} 
\left[\left(\frac{u-s-t}{2}\right)
\left(\frac{B-A}{2}\right) - 2 (I+J)
+\frac{\pi}{2}(u+s +2 m_c^2) + D\right] 
\label{dddmenosoff}
\end{equation}
for $D^*$ off-shell. Here $\lambda = \lambda(s,u,t) = 
s^2+t^2+u^2-2st-2su-2tu$, $s=p^2$, $u=p'^2$, $t=-Q^2$ 
and $A$, $B$,  $D$, $I$ and $J$ are functions of $(s,t,u)$, given by 
the following expressions:
\begin{eqnarray}
A&=&\frac{2\pi}{\sqrt{s}}\left( \overline{k_0}
- \frac{\overline{|\vec k|}p'_0}{|\vec p'|} 
\cos\overline{\theta}\right) \label{M}; \;\;\;\;\;\;\;\;\;\;\;\;\;
B= 2\pi\frac{\overline{|\vec k|}}{ |\vec p'|} \cos\overline{\theta} \,\,
\label{N} ; \;\;\;\;\;\;\;\;\;\;\;\;\;
D=-\pi \overline{|\vec k|}^2 
\left(1 - \cos^2\overline{\theta} \right) \label{G}; \\ 
I&=&\frac{\pi{\overline{|\vec k|}}^2}{\sqrt{s}}
\left(1 - \cos^2\overline{\theta}\right)
\left(\frac{\overline{|\vec k|} p'_0}{|\vec p'|}
\cos \overline{\theta}-\overline{k_0}\right)
\label{C} ; \;\;\;\;\;\;\;\;\;\;\;\;\;
J=-\frac{\pi\overline{|\vec k|}^3 }{|\vec p'|} 
\left(1 - \cos^2\overline{\theta}\right) \cos\overline{\theta} \,\, \label{D};  \\
\end{eqnarray}
where 
\begin{eqnarray}
p'_0&=&\frac{s+u-t}{2\sqrt{s}}    \label{pl0}; \;\;\;\;\;\;\;
|\vec p'|^2=\frac{\lambda}{4s}     \label{vpl};\;\;\;\;\;\;\
\overline{k_0}= \frac{s-m^2}{2\sqrt{s}}   \label{k0b} ;\;\;\;\;\;\;\\
\overline{|\vec k|}^2&=& \overline{k_0}^2-m^2   \label{vk}; \;\;\;\;\;\;
\cos\overline{\theta}=-\frac{u+\eta m_c^2-2p'_0\overline{k_0}}
{2|\vec p'|\overline{|\vec k|}}   \label{ctheta}; 
\end{eqnarray}
with  $\eta=1$ for $\rho$ off-shell and $\eta=-1$ 
for $D^*$ off-shell.

\begin{figure}[h]
\begin{picture}(6,3.5)
\put(0,0.5){\vector(1,0){1.5}}
\put(1.5,0.5){\line(1,0){0.75}}
\put(2.75,0.5){\line(1,0){0.75}}
\put(2.25,0.5){\circle*{0.15}}
\put(2.75,0.5){\circle*{0.15}}
\put(3.5,0.5){\vector(1,0){1.5}}
\put(1.5,0.5){\vector(1,1){1}}
\put(2.5,1.5){\vector(1,-1){1}}
\put(2.5,3){\vector(0,-1){1.5}}
\put(2.65,2.75){$q_\alpha$}
\put(0.25,0.65){$p_\mu$}
\put(4.55,0.65){$p'_\nu$}
\put(2.225,0.17){$\langle \bar u u\rangle$}
\put(1.85,1.1){$d$}
\put(3,1.1){$c$}
\put(2.4,1.2){$y$}
\put(1.75,0.53){$0$}
\put(3.05,0.53){$x$}
\put(1.75,2.2){$D*$}
\put(0.35,0.1){$\rho$}
\put(4,0.1){$D*$}
\end{picture}
\caption{Contribution of the $u\bar u$ condensate to the $D^*$ off-shell
correlator.}
\label{fig2}
\end{figure}
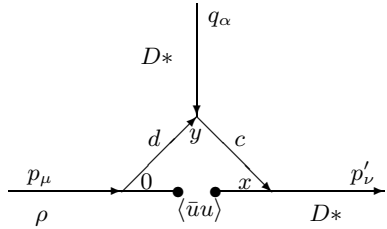

The contribution of the quark condensate which survives after the
double Borel transform is represented in Fig.~\ref{fig2} for the 
$D^*$ off-shell case and is given by 
\begin{equation}
\Gamma_c^{(D^*)} = -\frac{m_c \langle \bar q q\rangle}
{p^2 (p'^2 -m_c^2)} 
\end{equation}
where $\langle \bar q q\rangle$ is the light quark condensate. For the
$\rho$ off-shell there is no quark condensate contribuition.
 
We  expect  the perturbative contribution to dominate  the 
OPE, because we are dealing with heavy quarks. For this reason, we do not 
include the gluon and quark-gluon condensates in the present work.

The resulting vertex functions in the QCD side for the structure 
$g_{\mu\alpha} (q )_\nu$ are written as
\begin{equation}
\Gamma^{(\rho)}(p,\pli)=
-\frac{1}{4\pi^2}\int_{m_c^2}^{s_0} ds
\int_{m_c^2+t}^{u_0} du \:
\frac{\rho^{(\rho)}(s,u,Q^2)}{(s-p^2)(u-{\pli}^2)} \label{qcdrho}
\end{equation}
for $\rho$ off-shell and
\begin{equation}
\Gamma^{(D^*)}(p,\pli)=
-\frac{1}{4\pi^2}\int_{0}^{s_0} ds
\int_{t}^{u_0} du \:
\frac{\rho^{(D^*)}(s,u,Q^2)}{(s-p^2)(u-{\pli}^2)}+
\Gamma_c^{(D^*)} \label{qcdjp}
\end{equation}
for $D^*$ off-shell, where, as usual, we have already transferred the 
continuum contribution from the hadronic side to the QCD side, through 
the introduction of the continuum thresholds $s_0$ and $u_0$ \cite{io2}.

\subsection{The phenomenological side}

The $\rho D^* D^*$  vertex can be studied with hadronic degress of freedom. The 
corresponding three-point functions,  Eqs.~(\ref{correrhooff}) and 
(\ref{corredmenosoff}), will be written in terms of hadron masses, decay constants and
form factors. This is the so called phenomenological side of the sum rule and it 
is based on the interactions at the hadronic level, which  are described here by the   
following effective Lagrangian  \cite{linko}
\beqa
&&\mathcal{L}_{\rho D^{*}D^{*}}=
ig_{\rho D^{*}D^{*}} \Big [  \Big ((\partial_{\mu}D^{*\nu})
\vec{\tau} . \bar{D^{*}_{\nu}}
-D^{*\nu}\vec{\tau} .\partial_{\mu}\bar{D^{*}_{\nu}} \Big ) . \vec{\rho^{\mu}}
+ \Big (D^{*\nu} \vec{\tau}. \partial_{\mu} \vec{\rho_{\nu}}
\nonumber\\
&&- \partial_{\mu} D^{*\nu}\vec{\tau}. \vec{\rho_{\mu}} \Big)\bar{D^{*\mu}}
+D^{*\mu}\Big(\vec{\tau}.\vec{\rho^{\nu}}\partial_{\mu}\bar{D^{*}_{\nu}}
-(\vec{\tau}.\partial_{\mu}\vec{\rho^{\nu}})\bar{D^{*}_{\nu}} \Big) \Big],
\label{lagr}
\eeqa
from where one can extract the matrix element associated with the
$\rho D^*D^*$ vertex.
The meson decay constants, $f_{\rho}$ and $f_{D^*}$, are
defined by the following  matrix elements:
\beq
\langle 0|j^{\mu}_{\rho}|{\rho(p)}\rangle= m_{\rho} f_{\rho} \epsilon^{\mu}_{\rho}
(p)
\label{frho}
\eeq
and
\beq
\langle 0|j_{\nu}^{D^{*}}|{D^{*}(p)}\rangle= m_{D^{*}} f_{D^{*}} 
\epsilon^{\nu*}_{D^{*}}(p) \, ,
\label{fd}
\eeq
where $\epsilon^{\mu}_{\rho}$ and $\epsilon^{\nu}_{D^{*}}$ are the
polarization vectors of the  $\rho$ and $D^{*}$ mesons respectively. 
Saturating Eqs.~(\ref{correrhooff}) and (\ref{corredmenosoff}) with the
$\rho $ and two $D^*$ states and using Eqs.~(\ref{frho})
and (\ref{fd}) we arrive at
\beqa
&&\Gamma_{\mu \nu\alpha }^{(\rho)}= -g^{(\rho)}_{\rho D^*D^*}(Q^2) \sqrt{2}
\frac{f^2_{D^*}f_{\rho}m^2_{D^*}m_{\rho}}
{(P^2+m^2_{D^*})(Q^2+m^2_{\rho})({P^\prime}^2 +m^2_{D^*})}
\left(-g_{\mu\muli}+{p_\mu p_{\muli}\over m^2_{D^*}}\right)\times\nonumber\\
&&\left(-g_{\nu\nuli}+{\pli_\nu \pli_{\nuli}\over m^2_{D^*}}\right)
\left(-g_{\alpha\ali}+{q_\alpha q_{\ali}\over m^2_{\rho}}\right)
\left[(p+\pli)^{\ali}g^{\muli\nuli}+(2 \pli-p)^{\nuli}g^{\ali\muli}-
(2 p-\pli)^{\muli}g^{\ali\nuli}\right],
\label{allstru}
\eeqa
when the $\rho$ is off-shell, with a similar expression for the
$D^*$ off-shell: 
\beqa
&&\Gamma_{\mu \nu\alpha }^{(D^*)}= -g^{(D^*)}_{\rho D^*D^*}(Q^2) \sqrt{2}
\frac{f^2_{D^*}f_{\rho}m^2_{D^*}m_{\rho}}
{(P^2+m^2_{\rho})(Q^2+m^2_{D^*})({P^\prime}^2 +m^2_{D^*})}
\left(-g_{\mu\muli}+{p_\mu p_{\muli}\over m^2_{\rho}}\right)\times\nonumber\\
&&\left(-g_{\nu\nuli}+{\pli_\nu \pli_{\nuli}\over m^2_{D^*}}\right)
\left(-g_{\alpha\ali}+{q_\alpha q_{\ali}\over m^2_{D^*}}\right)
\left[(p+\pli)^{\ali}g^{\muli\nuli}-(2 p  - \pli)^{\nuli}g^{\ali\muli} -
(2 \pli - p)^{\muli}g^{\ali\nuli}\right],
\label{phendsoff}
\eeqa
The contractions of $\muli,~\nuli$ and $\ali$ in the above 
equation will lead to the fourteen Lorentz structures appearing in 
Eq.~(\ref{trace}).
We can see from Eq.~(\ref{allstru}) that the form factor
$g^{(\rho)}_{\rho D^*D^*}(Q^2)$ is the same for all the structures
and thus can be extracted from sum rules written for any of these
structures. The resulting phenomenological invariant amplitudes associated with the 
structure $g_{\alpha \mu} (q)_{\nu}$ are
\begin{equation}
\Gamma^{(\rho)}_{ph}(p^2,{\pli}^2,Q^2)= g^{(\rho)}_{\rho D^*D^*}(Q^2)
\frac{\sqrt{2} f_{D^*}^2f_{\rho}m^2_{D^*}m_{\rho} (2-\frac{m^2_{\rho}}{2 m^2_{D^*}}) }
{(P^2+m^2_{D^*})(Q^2+m^2_{\rho})(P^{\prime2} +m^2_{D^*})}
\label{phenrhooff}
\end{equation}
for the $\rho$ off-shell, and 
\begin{equation}
\Gamma^{(D^*)}_{ph}(p^2,{\pli}^2,Q^2)= g^{(D^*)}_{\rho D^*D^*}(Q^2)
\frac{f^2_{D^*}f_{\rho}m^2_{D^*}m_{\rho}(Q^2 +4 m^2_{D^*})   }
{(P^2+m^2_{D^*})(Q^2+m^2_{D^*})(P^{\prime2} +m^2_{\rho}) \sqrt{2}m^2_{D^*}}
\label{phendmenosoff}
\end{equation}
for ${D^*}$ off-shell.

In order to improve the matching between the two sides of the sum rules
we perform a double Borel transformation \cite{io2} in the variables 
$P^2=-p^2\rightarrow M^2$ and $P'^2=-{\pli}^2\rightarrow M'^2$, 
on both invariant amplitudes $\Gamma$ and $\Gamma_{ph}$. 
Equating the results we get the final expressions for the sum rules  
which allow us to obtain the form factors 
$g^{(T)}_{\rho D^*D^*}(Q^2)$ appearing in 
Eqs.~(\ref{phenrhooff})--(\ref{phendmenosoff}), where $T$ is $\rho$ or  
$D^{*}$. 
In this work we use the following relations between the Borel masses $M^2$ and 
$M'^2$: $\frac{M^2}{M'^2} = \frac{m^2_{\rho}}{m^2_{D^*}}$
for a  $D^*$ off-shell  and $M^2 = M'^2$ for a $\rho$ off-shell.

\begin{table}[h]
\begin{tabular}{|c|c|c|c|c|c|} \hline
$m_c (\GeV)$ & $m_{D^*} (\GeV)$ & $m_{\rho} (\GeV)$ & $f_{D^*} (\GeV)$ & 
$f_{\rho} (\GeV)$ & 
$\langle \bar q q \rangle (\GeV)^3$  \\ \hline 
1.35&2.01&0.778&0.240& 0.161 &$(-0.23)^3$ \\ \hline
\end{tabular}
\caption{Parameters used in the calculation.}
\label{tableparam}
\end{table}

\section{Results and discussion}

Table \ref{tableparam} shows the values of the parameters used in the present 
calculation. We used the experimental value for  $f_{\rho}= \frac{m_{\rho}}{g_{\rho}}$, 
with $g_{\rho}=4.79 $ \cite{pdg}  , and took $f_{D^*}$ from ref.~\cite{khod}.
The continuum thresholds are given by 
$s_0=(m+ \Delta_s)^2$ and $u_0=(m+\Delta_u)^2$, where $m$ 
is the mass of the incoming meson. 
\begin{figure}[ht] 
\centerline{\epsfig{figure=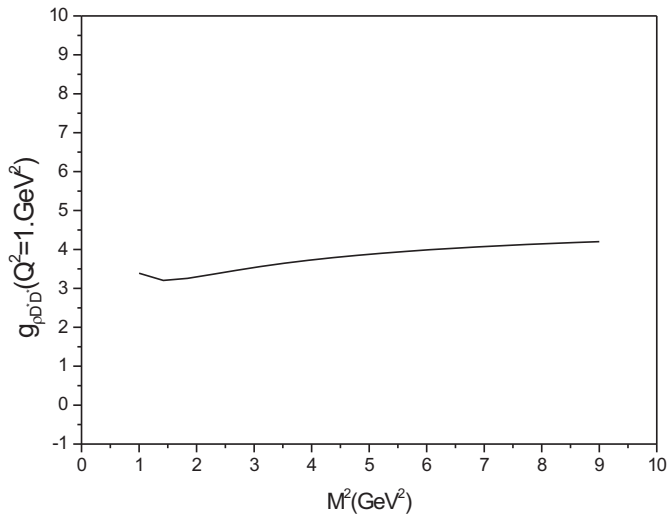,height=80mm}}
\caption{$g^{(\rho)}_{\rho D^*D^*}(Q^2=1.0 \GeV^2)$ as a function 
of the Borel mass $M^2$.}
\label{fig3}
\end{figure}
Using $\Delta_s=\Delta_u = 0.5 \GeV $ for the continuum thresholds 
and fixing $Q^2=1 \GeV^2$, we found a good stability of the 
sum rule for $g_{\rho D^* D^*}^{(\rho)}$ for $M^2$ in the interval 
$ 1 < M^2 < 9 \GeV^2$, as can be seen in Fig.~\ref{fig3}. Within this interval 
we need to choose the best value of the  Borel mass to 
extract the  coupling constant of the vertex. 
It is well known in QCDSR that if we choose a too small value of the Borel 
variable $M^2$, then the sum rule will be dominated by the pole, but the 
convergence of the OPE is poor. On the other hand, if $M^2$ is too large, 
then the OPE convergence is good but the sum rule is dominated by the 
continuum. The best value of the Borel mass is the one with which both criteria are  
reasonably satisfied.

\begin{figure}[ht] 
\centerline{\epsfig{figure=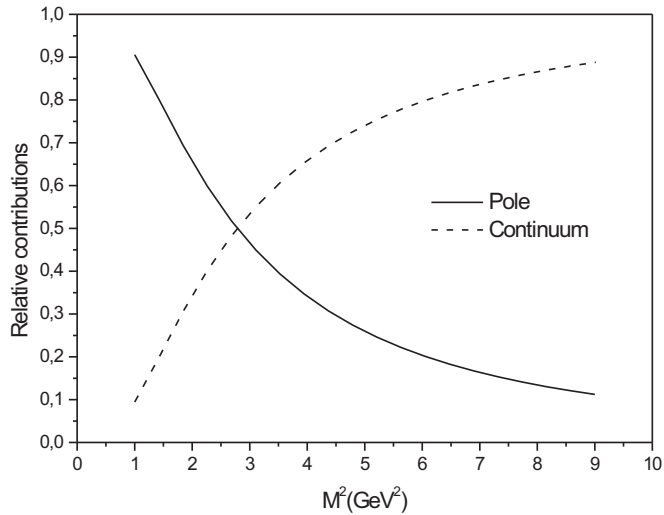,height=80mm}}
\caption{Pole (solid line) and continuum (dashed line) contribuition to  
$g^{(\rho)}_{\rho D^*D^*}(Q^2=1\GeV^2, M^2)$, 
as a function of the Borel mass $M^2$.}
\label{fig4}
\end{figure}

In Fig.~\ref{fig4} we show   the pole contribution (solid line) 
and the continuum contribution (dashed line) divided by their sum 
 as a function of Borel mass. 
In the case of $\rho$ off-shell, we see that the pole contribution is bigger than the
continuum one in the Borel window $1 < M^2 < 3 \,\, GeV^2 $. 
The best Borel mass seems to be $ M^2= 2.0 \,\, GeV^2$. Also for this value of $M^2$ 
the coupling $\alpha_s$ is aproximately $0.2$ and this suggests that perturbative 
corrections are small.

\begin{figure}[ht] 
\centerline{\epsfig{figure=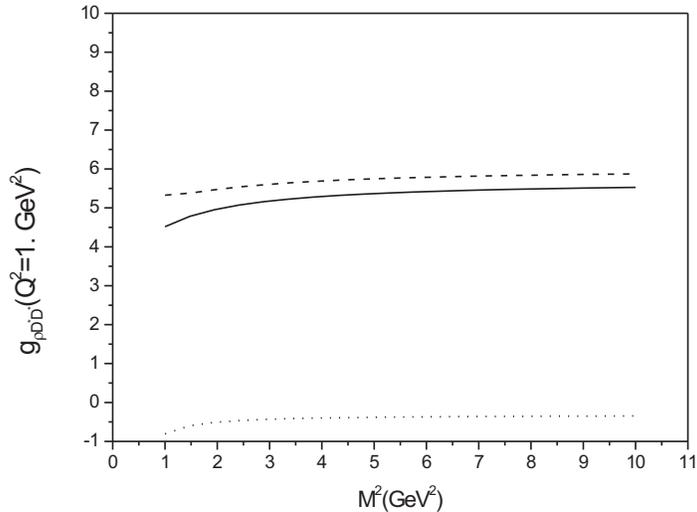,height=80mm}}
\caption{$g^{(D^*)}_{\rho D^*D^*}(Q^2=1\GeV ^2)$ as a
function of the Borel mass. We show the perturbative contribution (dashed line), 
quark condensate contribution (dotted line) and total (solid line).}
\label{fig5}
\end{figure}

\begin{figure}[ht] 
\centerline{\epsfig{figure=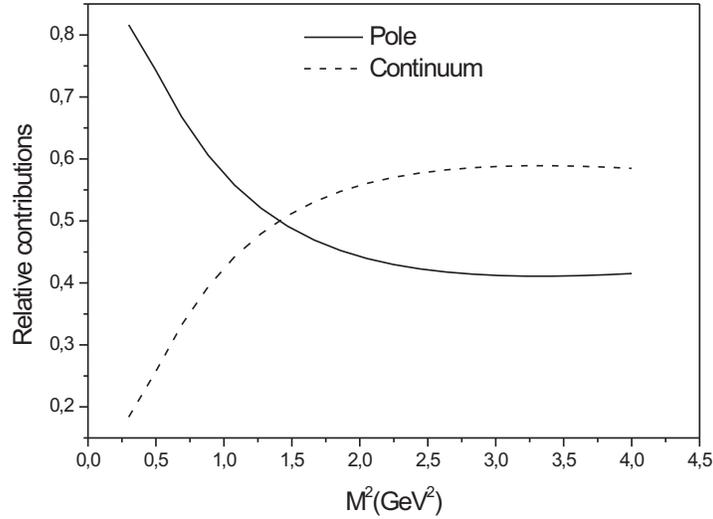,height=80mm}}
\caption{Pole versus continuum contributions  to  $g^{(D^*)}_{\rho D^*D^*}(Q^2=1\GeV^2)$  
as a function of the Borel mass $M^2$.}
\label{fig6}
\end{figure}

In the case of $g_{\rho D^* D^*}^{(D^*)}$ the interval for stability is 
$1 < M^2 < 10 \, \GeV^2 $, as can be seen in Fig.~\ref{fig5}.
In order to choose the Borel mass we proceed in the same way as before and  
we analyse  the pole and continuum contributions. As indicated  in
Fig.\ref{fig6} in  the window  $ 0.5 < M^2 < 1.5 \, \GeV^2 $ the pole contribution 
dominates. We choose  $M^2 = 1.5 \, GeV^2 $.

\begin{figure}[ht] 
\centerline{\epsfig{figure=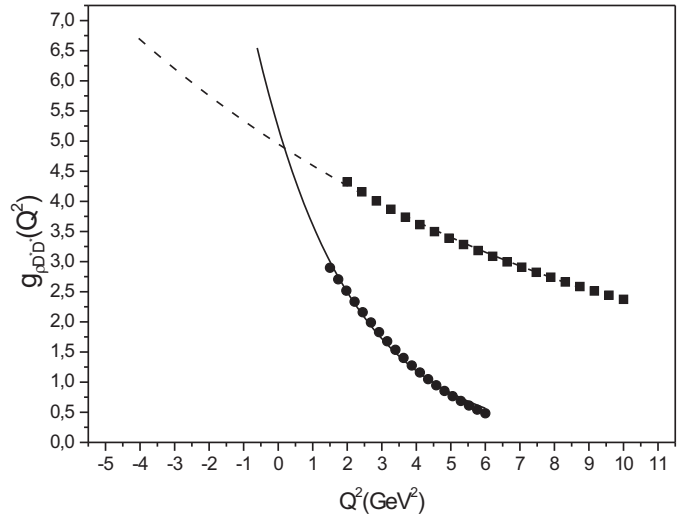,height=80mm}}
\caption{$g^{(\rho)}_{\rho D^*D^*}$ (circles) and 
$g^{(D^*)}_{\rho D^*D^*}$ (squares) QCDSR form factors as a function of
$Q^2$. The solid and dashed lines 
correspond to the  exponential parametrizations of the QCDSR data with the 
two forms mentioned in the text.}
\label{fig7}
\end{figure}

Having determined $M^2$ we calculated the $Q^2$ dependence of the form factors. 
We present the results in Fig.~\ref{fig7}, where the circles correspond to the 
$g_{\rho D^* D^*}^{(\rho)}(Q^2)$ form factor in the  interval where the sum rule is valid. 
The squares are the result of the sum rule for the $g_{\rho D^* D^*}^{(D^*)}(Q^2)$ 
form factor. 

\vspace{0.5cm}
In the case of an off-shell $\rho$ meson, our numerical results can be 
fitted by the following exponential parametrization (shown by the dotted line in
Fig.~\ref{fig7}):
\begin{equation}
g_{\rho D^* D^*}^{(\rho)}(Q^2)= 5.22 e^{-Q^2/2.70} \;.
\label{exprho}
\end{equation}
As in our previous works \cite{bclnn01,mnns02,smnn04,mnns05,cdnn05,bcnn05}, 
we define the coupling constant as the value of the 
form factor at $Q^2= -m^2_{m}$, where $m_{m}$ is the mass of the off-shell meson. 
Therefore, using $Q^2=-m_{\rho}^2$ in Eq~(\ref{exprho}), the resulting coupling 
constant is:
\begin{equation}
g_{\rho D^* D^*}= 6.55 \;.  \label{couplingrho}
\end{equation}

For an off-shell $D^{*}$ meson, our sum rule results  can also be 
fitted by an exponential parametrization, which is represented by the
dashed line in Fig.~\ref{fig7}:  
\begin{equation}
g_{\rho D^* D^*}^{(D^{*})}(Q^2)= 4.95 e^{-Q^2/13.33}\;.
\label{expdmenos}
\end{equation}
Using $Q^2=-m_{D^{*}}^2$ in Eq~(\ref{expdmenos}) we get:
\begin{equation}
g_{\rho D^* D^*}= 6.70,
\label{couplingdmenos}
\end{equation}
in a good agreement with the result of Eq.(\ref{couplingrho}).

\begin{figure}[ht] 
\centerline{\epsfig{figure=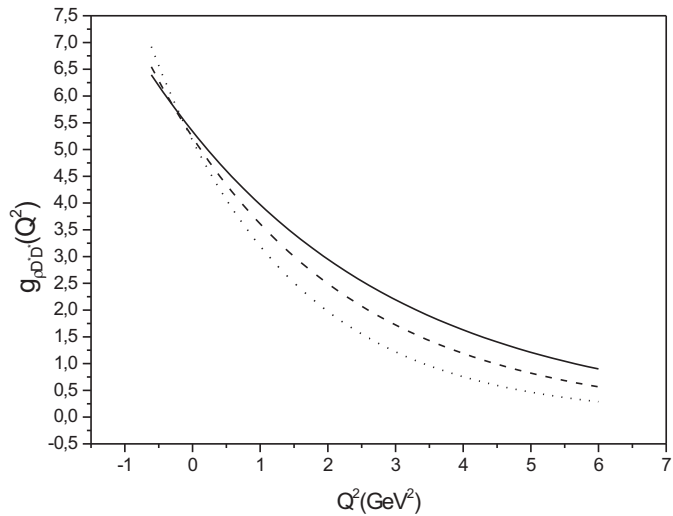,height=80mm}}
\caption{Dependence of the form factor on the continuum threshold for
the $\rho$ off-shell case. The dashed curve  corresponds to 
$\Delta_{s,u} = 0.5\GeV$, the solid one  to $\Delta_{s,u}= 0.6\GeV$
and the dotted curve to $\Delta_{s,u}= 0.4\GeV$.}
\label{fig8}
\end{figure}

In order to study the dependence of our results with the continuum
threshold, we vary $\Delta_{s,u}$ between 
$0.4\GeV\le \Delta_{s,u}\le 0.6\GeV$ in the parametrization
corresponding to the case of an  off-shell  $\rho$. As can be seen in 
Fig.~\ref{fig8}, this procedure gives us an uncertainty interval of 
$6.40  \le g_{\rho D^* D^*} \le 6.92 $ for the coupling constant.

Concluding, the two cases considered here, off-shell $\rho$ or $D^*$,
give compatible results for the coupling constant, evaluated using the QCDSR 
approach. Considering the uncertainties in the continuum
thresholds we obtain:
\begin{equation}
g_{\rho D^* D^*}= 6.6 \pm0.31\;. \label{finalcoupling}
\end{equation}
This  generic value of the coupling constant can be easily related to the couplings 
of the specific charge states. From Eq. (\ref{lagr}) we arrive at:
\begin{equation}
g_{\rho D^* D^*}= \frac{g_{\rho^- D^{*0} D^{*+}}}{\sqrt{2}} = 
 \frac{g_{\rho^+ D^{*0} D^{*+}}}{\sqrt{2}} = -  g_{\rho^0 D^{*+} D^{*+}}
=  g_{\rho^0 D^{*0} D^{*0}}
\end{equation}

From  Eqs.~(\ref{exprho}) and (\ref{expdmenos}) we can 
also extract the cut-off parameter, $\Lambda$, associated with the form factors.
We get $\Lambda \sim 1.64\GeV$ for an off-shell $\rho$ meson and  $\Lambda\sim
3.65 \GeV$ for an off-shell $D^*$. The cut-off values obtained here follow the same
trend as observed in Refs \cite{bclnn01,mnns02,smnn04,mnns05}  : 
the value of the cut-off is directly associated with the mass of the off-shell 
meson probing the vertex. The form factor is harder if the off-shell meson 
is heavier.

As for the value of  $g_{\rho D^* D^*}$, this coupling has not been discussed in the 
literature  as much as those involving the $J/\psi$ and there are only few works 
presenting estimates for it.  The starting point in these estimates is always the 
SU(4) symmetry. According to SU(4) we should expect:
\beq
g_{J/\psi  D^* D^*} = g_{J/\psi D D}
\label{su41}
\eeq
\beq
g_{\rho D^* D^*} = g_{\rho D D}
\label{su42}
\eeq
and
\beq
g_{\rho D^* D^*} = \frac{\sqrt{6}}{4} g_{J/\psi  D^* D^*}
\label{su43}
\eeq

From our previous works \cite{mnns02,bcnn05}
we find that in QCDSR Eq. (\ref{su41}) is satisfied. However, from \cite{bclnn01} and 
from the present work we conclude that Eq. (\ref{su42}) is not satisfied, 
since $g_{\rho D^* D^*}= 6.6 \pm0.26 $ 
whereas  $g_{\rho D D}= 3.04 \pm 0.30$ 
\footnote{Due to a difference in definitions, this 
number is a factor $\sqrt{2}$ smaller than the one quoted in \cite{bclnn01}.}. 
Eq. (\ref{su43}) is not true either because  
$g_{J/\psi  D^* D^*} = 6.2 \pm 0.9$. These relations are violated at the level of 
50 \%. This is not surprising since the
mass difference starts to play an important role when we go from the  heavier 
vector mesons to $\rho$.  As for the absolute value, the existing estimates, used
in \cite{linko,su}, lead to  $g_{\rho D^* D^*}= 2.52 $. Our result is a factor two 
larger.

\acknowledgments
This work has been supported by CNPq, CAPES and FAPESP.

\end{document}